\documentclass[12pt,preprint]{aastex}
\usepackage{graphicx,subfigure,amsmath}
\providecommand{\e}[1]{\ensuremath{\times 10^{#1}}}
\newcommand\gsim{\,\lower3pt\hbox{$\sim$}\llap{\raise2pt\hbox{$>$}}\,}
\newcommand\lsim{\,\lower3pt\hbox{$\sim$}\llap{\raise2pt\hbox{$<$}}\,}

\begin{document}
\title{The Rise of Active Region Flux Tubes in the Turbulent Solar Convective Envelope}

\author{Maria A. Weber\altaffilmark{1}$^{,}$\altaffilmark{2}, Yuhong Fan\altaffilmark{1}, and Mark S. Miesch\altaffilmark{1}}
\affil{High Altitude Observatory, National Center for Atmospheric
Research\altaffilmark{1}, 3080 Center Green Drive, Boulder, CO 80301}

\altaffiltext{1} {The National Center for Atmospheric Research
is sponsored by the National Science Foundation}
\altaffiltext{2}{Graduate Student in the Department of Physics, Colorado State University, Fort Collins, CO 80523, USA}

\begin{abstract}
We use a thin flux tube model in a rotating spherical shell of turbulent convective flows to study how active region scale flux tubes rise buoyantly from the bottom of the convection zone to near the solar surface.  We investigate toroidal flux tubes at the base of the convection zone with field strengths ranging from 15 kG to 100 kG at initial latitudes ranging from 1$^{\circ}$ to 40$^{\circ}$ with a total flux of $10^{22}$ Mx.  We find that the dynamic evolution of the flux tube changes from convection dominated to magnetic buoyancy dominated as the initial field strength increases from 15 kG to 100 kG. At 100 kG, the development of $\Omega$-shaped rising loops is mainly controlled by the growth of the magnetic buoyancy instability.  However, at low field strengths of 15 kG, the development of rising $\Omega$-shaped loops is largely controlled by convective flows, and properties of the emerging loops are significantly changed compared to previous results in the absence of convection. With convection, rise times are drastically reduced (from years to a few months), loops are able to emerge at low latitudes, and tilt angles of emerging loops are consistent with Joy's Law for initial field strengths of $\gsim 40$ kG.  We also examine other asymmetries that develop between the leading and following legs of the emerging loops.  Taking all the results together, we find that mid-range field strengths of $\sim$ 40 - 50 kG produce emerging loops that best match the observed properties of solar active regions.
\end{abstract}

%-----------------------------------------------------------------------------------------------------------------------------------

\section{Introduction}

The Sun's cyclic large scale magnetic field with a period of 22 years is believed to be sustained by a dynamo mechanism.  The Hale polarity law \citep{hale19} of solar active regions indicates the presence of a large scale subsurface toroidal magnetic field generated by the solar dynamo.  The current prevailing picture is that the large scale toroidal magnetic field responsible for the formation of solar active regions is amplified and stored at the base of the solar convection zone \citep[e.g.][]{gilman00,char10}.  Thus answering the question of how active region flux tubes rise from the base of the convection zone to the solar surface is vitally important in the development of solar dynamo theory.  

A large body of calculations based on a so called thin flux tube approximation has provided important insights into the dynamic evolution of $\Omega$-shaped rising loops in the solar convective envelope \citep[see review by e.g.][]{fan2009}. The thin flux tube model assumes that the radius of the tube is significantly smaller than all other relevant scales of variation such as the pressure scale height and the curvature of the tube. Given the observed scale of flux $\sim 10^{22}$ Mx in a large solar active region, and assuming that the field strength of the flux tubes at the base of the convection zone is at least 10 kG, which is the order of field strength in equipartition with convection, one finds that the above condition for the thin flux tube approximation is satisfied in the bulk of the solar convection zone. The thin flux tube model solves for the mean motion of each tube segment under the influence of various integrated forces acting on the tube segment \citep[see e.g.][]{spruit1981,longcope_klapper1997}. Results from thin flux tube models without the influence of convective flows suggest that the field strength of the toroidal magnetic field at the base of the solar convection zone needs to be in the range of about 30 kG to about 100 kG in order for the latitude of emergence and the tilt angles of the emerging loops to be consistent with the observed properties of solar active regions.  If the field strength is $\leq$ 20 kG , the poleward deflections of the trajectories of the rising flux tubes by the Coriolis force are too large such that the emerging latitudes are inconsistent with the observed sunspot latitudes \citep[e.g.][]{cali95}.

The thin flux tube calculations show that the Coriolis force acting on the rising $\Omega$-loops produces asymmetries between the leading and the following legs of the loop (the leading leg is the proceeding portion of the loop in the direction of solar rotation), which provide explanations for several observed asymmetries between the leading and following polarities of bipolar active regions. First, a slight tilt of the emerging loop is produced, with the leading leg being closer to the equator than the following, consistent with the observed Joy's law of active region tilt angles \citep[e.g.][]{dsilva93}. Second, an asymmetric geometric shape of the emerging loop is produced, where the leading side is inclined at a smaller angle to the surface \citep[e.g.][]{cali95}. This may give rise to an apparent faster proper motion of the leading polarity as the loop emerges through the surface, which is observed \citep{van_1990}. A third asymmetry is in the field strength of the loop, with a stronger field strength in the leading side compared to the following, which may explain the observed more coherent morphology of the leading polarity of an active region \citep[e.g.][]{fan93}. However, this field asymmetry is found to depend on the assumed initial condition of the flux tube \citep{cali98,fan_fisher1996}.
 
The studies discussed previously do not address the dynamical effects of turbulent convection on a rising flux tube.  These are shown to be adequate simplifications when the flux tube's magnetic field strength $B \gtrsim (H_{p}/a)^{1/2}B_{eq} \sim 3 - 5 B_{eq}$, where $B_{eq} \sim$ 10 kG is the field strength for which the magnetic energy density is in equipartition with the kinetic energy density of convection, $H_{p}$ is the local pressure scale height, and $a$ is the tube radius \citep{fan03}.  Solar cycle dynamo models which take into account the dynamic effects of the Lorentz force from the large-scale mean fields suggest that the toroidal magnetic field generated at the base of the convection zone is $\sim$ 15 kG, and most likely cannot exceed 30 kG \citep[e.g.][]{rempela,rempelb}.  Therefore, it is important to understand how toroidal flux tubes in the range of 15 kG to 30 kG rise through a turbulent solar convective envelope.   Thin flux tube simulations which neglect the effect of convection \citep[e.g. see review by][]{fan2009}, show that tubes in this range of field strengths tend to be deflected significantly poleward by the Coriolis force during their rise, and so have difficulty reproducing the emergence of active regions at low latitudes, as well as Joy's law.  At these field strengths, convective downdrafts are capable of pinning down the flux tube while helical upflows between the downdrafts can boost the rise of the flux tube, and therefore the resulting emerging loops would be significantly different from those produced in the absence of convection.

A few studies have been performed to investigate the the buoyant rise of a flux tube in a turbulent convective velocity field \citep[e.g.][]{fan03,jouve09}. \citet{fan03} carried out 3D Magnetohydrodynamic (MHD) simulations of the evolution of a buoyant magnetic flux tube in a stratified convective velocity field in a local Cartesian geometry without the effect of solar rotation. It is found that buoyant flux tubes with $B \leq (H_{p}/a)^{1/2}B_{eq} \sim 30 $ kG are strongly influenced by convection, where portions of the tubes in convective downdrafts becomed pinned down to the bottom of the domain, while the rise speed of sections within upflow regions is significantly boosted. It is found that the evolution of the flux tube is no longer sensitive to the twist of the tube in this convection dominated regime. \citet{jouve09} have carried out the first set of global anelastic MHD simulations of the buoyant rise of an initially toroidal flux ring in a rotating, fully convective spherical shell, possessing self-consistently generated mean flows such as meridional circulations and differential rotation, representative of the conditions of the solar convective envelope. Due to the limited numerical resolution and the relatively high magnetic diffusivity in these global 3D simulations, flux tubes with a very large initial field strength (ranging from 150 kG to 600 kG) and a large radius, corresponding to a total flux on the order of a few times $10^{23}$ Mx, significantly greater than the typical active region fluxes, are considered.   These large values are needed such that the rise times of flux tubes are less than their diffusive time scales. Because most of the cases considered are essentially in the magnetic buoyancy dominated regime, the rising toroidal flux tube only develops rather moderate undulations by the influence of the convective flows, and $\Omega$-shaped tubes with undulations extending the depth of the convection zone are not found. 

In this study, we use the thin flux tube model, with the inclusion of a (separately computed) 3D turbulent convective velocity field in a rotating model solar convective envelope, to study its effects via the aerodynamic drag force, together with the forces of magnetic buoyancy, tension, and the Coriolis force, on the dynamic evolution of the emerging $\Omega$-loops. Since the thin flux tube model preserves the frozen-in magnetic field condition, it does not suffer the ÒerosionÓ of magnetic buoyancy and the tension force due to artificial diffusion of the magnetic field, as do the Eulerian based multi-dimensional simulations.  Although these thin flux tube models do not capture the internal structure and the possible fragmentation of the flux tube, they provide an initial step towards understanding the effects of convective flows in relation to other forces acting on the rising flux tubes. 

This paper is organized as follows. The equations of thin flux tube dynamics and the separately computed convective flow are discussed in $\S$ \ref{sec:model}.  Results of our simulations are presented in $\S$ \ref{sec:results}, focusing in particular on the overall influence of convection on the flux tubes,  the latitude of emergence and tilt angles of the flux tube, the asymmetry in field strength and geometry of the emerging flux loops, and the rise time for the flux tubes to reach the surface.  We perform these thin flux tube simulations for a range of magnetic fields (15 kG to 100 kG) from near-equipartition to super-equipartition initial field strengths at varying latitudes from 1$^{\circ}$ to 40$^{\circ}$ in both the northern and southern hemispheres.  A conclusion and discussion of the results is given in $\S$ \ref{sec:discuss}.

%------------------------------------------------------------------------------------------------------------------------------------

\section{Model Description}
\label{sec:model}

A discussion of the thin flux tube model, and how it is derived from the MHD equations under the assumption that the tube radius is thin compared to all other relevant scales of variation can be found in many previous publications \citep[see e.g.][]{spruit1981,longcope_klapper1997,fan2009}. For this work, the equations that describe the evolution of the thin flux tube are:
\begin{eqnarray}
\rho {d {\bf v} \over dt} & = & -2 \rho ( {\bf \Omega_0} \times {\bf v} )
-(\rho_e - \rho ) [{\bf g} - {\bf \Omega_0} \times ({\bf \Omega_0} \times
{\bf r})] + {\bf l} {\partial \over \partial s} \left ( { B^2 \over 8 \pi}
\right ) + {B^2 \over 4 \pi} {\bf k} 
\nonumber \\
& & - C_d {\rho_e | ({\bf v }-{\bf v}_e)_{\perp} |
({\bf v}-{\bf v}_e)_{\perp} \over ( \pi \Phi / B )^{1/2} },
\label{eq:eqn_motion}
\end{eqnarray}
\begin{equation}
{d \over dt} \left ( {B \over \rho} \right ) = {B \over \rho} \left [
{\partial ({\bf v} \cdot {\bf l}) \over \partial s} - {\bf v} \cdot
{\bf k} \right ] ,
\label{eqn_cont_induc}
\end{equation}
\begin{equation}
{1 \over \rho} {d \rho \over dt} = {1 \over \gamma p} {dp \over dt} ,
\label{eqn_adiab}
\end{equation}
\begin{equation}
p = {\rho R T \over \mu} ,
\label{eqn_state}
\end{equation}
\begin{equation}
p + {B^2 \over 8 \pi} = p_e ,
\label{eqn_pbalance}
\end{equation}
where, ${\bf r}$, ${\bf v}$, $B$, $\rho$, $p$, $T$, which are functions of the time $t$ and arc length $s$ measured along the tube, denote respectively the position, velocity, magnetic field strength, gas density, pressure, and temperature of a Lagrangian tube segment, ${\bf l} \equiv \partial {\bf r} / \partial s$ is the unit vector tangential to the flux tube and ${\bf k} \equiv \partial^2 {\bf r} / \partial s^2 $ is the tube's curvature vector, subscript '${\perp}$' denotes the component perpendicular to the flux tube, $\Phi=10^{22}$ Mx is the constant total flux of the tube, $\rho_e$, $p_e$, $T_e$, and $\mu$, which are functions of depth only, are respectively the pressure, density, temperature, mean molecular weight of the surrounding external plasma, ${\bf g}$ is the gravitational acceleration and a function of depth, ${\bf \Omega_0}$ is the angular velocity of the reference frame co-rotating with the sun, with $\Omega_0$ set to $2.7 \times 10^{-6}$ rad/s in this calculation, $C_d$ is the drag coefficient, set to 1 in this calculation, $\gamma$ is the ratio of specific heats, and ${\bf v}_e ({\bf r}, t)$ (discussed below) is a time dependent velocity field (relative to the rotating frame of reference) that impacts the dynamics of the thin flux tube through the drag force term. In the above equations, we do not introduce an explicit magnetic diffusion or kinematic viscosity term.  However, we do introduce a drag force, which is the last term in Eq. \ref{eq:eqn_motion}.  This describes the interaction of the external fluid with the flux tube in a high Reynolds number regime \citep[e.g.][]{batchelor}. The thin flux tube is untwisted, and discretized with 800 uniformly spaced grid points along its arc length $s$.  The numerical method used to solve for the flux tube evolution as determined by the above set of equations has been described in detail in \citet{fan93}.  

\begin{figure}
\epsscale{0.7}
\plotone{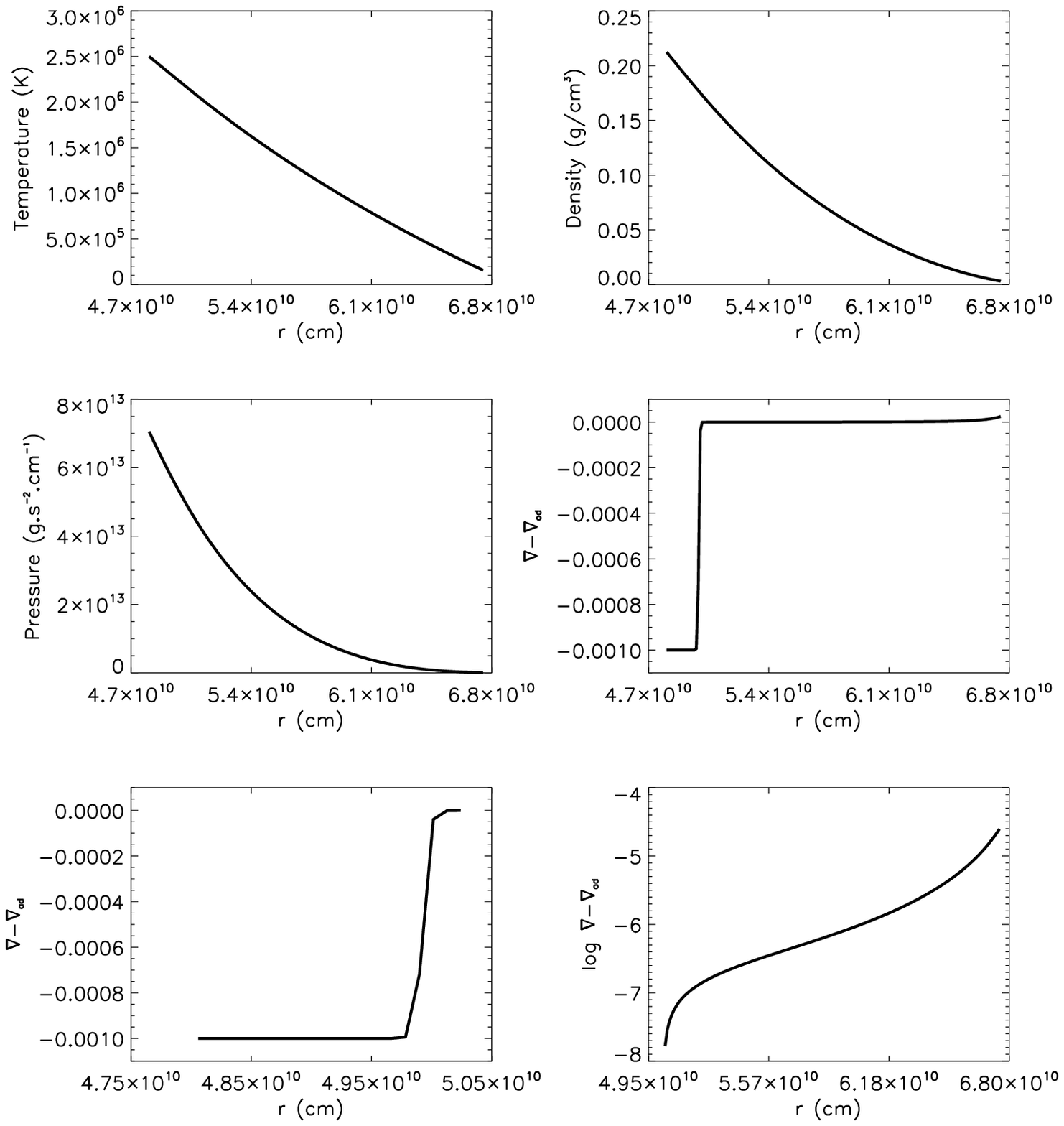}
\caption{Profiles of $T_e$, $\rho_e$, $p_e$ and adiabaticity factor $\delta$ for the entire simulation domain as a function of solar radius $r$ for the reference solar model.  The bottom left panel shows the sub-adiabaticity in the overshoot region, whereas the bottom right panel shows the logarithm of the super-adiabaticity in the convection zone.}
\label{fig:jcd}
\end{figure}

For the stratification of the external field free plasma, namely $\rho_e$, $p_e$, $T_e$, $\mu$, $g$, $\gamma$, and the super-adiabaticity, we use the reference solar model by \citet{jcd1996} for the solar convection zone with an extension of a simple polytropic, sub-adiabatically stratified thin overshoot layer, as described in \citet{fan_gong2000}.  Profiles of $T_e$, $\rho_e$, $p_e$, and the adiabaticity $\delta=\nabla - \nabla_{ad}$, where $\nabla=d$ $ln$ $T_{e}/d$ $ln$ $p_{e}$ and $\nabla_{ad}$ is the value of $\nabla$ one obtains by considering local adiabatic perturbations, are shown in Figure \ref{fig:jcd}.  The bottom left panel of Figure \ref{fig:jcd} shows the sub-adiabaticity of the thin overshoot layer, which extends from 4.8\e{10} cm to 5.026\e{10} cm.  The bottom right panel shows the logarithm of the super-adiabaticity in the convection zone, which extends from 5.026\e{10} cm to 6.75\e{10} cm.          

The main new ingredient of the current thin flux tube simulations is the inclusion into the drag force 
term of an external time-dependent convective velocity field ${\bf v}_e ({\bf r}, t)$ relative to the rotating frame of reference, computed separately from a 3-dimensional global convection simulation using the ASH (anelastic spherical harmonic) code, as described in \citet{mieschetal2006} (hereafter MBT06). In the anelastic approximation, convective flows and thermal variations are treated as a linear perturbation to a background state which is taken from a 1D solar structure model.  The computed ${\bf v}_e ({\bf r}, t)$ captures giant-cell convection and the associated mean flows (including the differential rotation and meridional circulation) in the rotating solar convective envelope spanning $r = 0.69 R_{\odot}$ to $r = 0.97 R_{\odot}$ (4.8\e{10} cm to 6.75\e{10} cm), resolved by a grid of 129 points in $r$, $256$ points in $\theta$, and $512$ points in $\phi$.  ASH is a pseudo-spectral code with horizontal and vertical basis functions given by spherical harmonics and Chebyshev polynomials, eached de-aliased by keeping only the lowest 2/3 of modes (maximum spherical harmonic degree $\ell_{max} = 170$ and Chebyshev degree $n_{max} = 86$). 

Simulation parameters and boundary conditions are similar to Case AB3 in MBT06.   In particular, the lower thermal boundary condition is the same as in Case AB3, with a latitudinal entropy gradient imposed in order to implicitly capture thermal coupling to the tachocline: $S(\theta,r_1) = C_P \left(a_2 Y_{20} + a_4 Y_{40}\right)$, where $S$ is the specific entropy per unit mass, $r_1$ is the inner boundary, $C_P$ is the specific heat at constant pressure, $Y_{\ell m}(\theta,\phi)$ is the spherical harmonic of degree $\ell$ and order $m$, $a_2 = 1.7\times 10^{-6}$, and $a_4 = -0.43 \times 10^{-6}$.  This helps promote a conical rotation profile but the strong rotational shear is attributed to the convective Reynolds stress rather than baroclinicity (see MBT06).   However, the radial entropy gradient imposed at the outer boundary is steeper, more in line with solar structure models (e.g. \citet{jcd1996});  $\partial S/\partial r = - 10^{-5}$ erg g$^{-1}$ K$^{-1}$ cm$^{-1}$ in this case compared to $-10^{-7}$ erg g$^{-1}$ K$^{-1}$ cm$^{-1}$ in Case AB3.   This implies a higher Rayleigh number, which is partially offset by higher values of the turbulent viscosity $\nu$ and thermal diffusivity $\kappa$.  In this simulation the values of $\nu$ and $\kappa$ at the outer boundary ($r=0.97R$) are $2\times10^{13}$ and $4\times 10^{13}$ cm$^2$ s$^{-1}$ respectively and each decreases with depth in proportion to the inverse square root of the background density $\hat{\rho}^{-1/2}$.  The density contrast across the domain is 69, which corresponds to 4.2 density scale heights (somewhat larger than AB3, due to a slightly deeper simulation domain).    This yields a mid-convection zone Rayleigh number $R_a$ of $5\times 10^6$ and Reynolds number $R_e$ of order 50.  The Rayleigh number is defined here as $R_a = g r^2 d \Delta S / (\nu \kappa C_P)$ where $g$ is the gravitational acceleration and $d = r_2-r_1$ is the depth of the layer.  The Reynolds number is given by $V_{rms} d / \nu$ where $V_{rms}$ is the root mean square velocity relative to the rotating reference frame..  This value of $R_a$ is somewhat larger than that used in the flux tube simulations of \citet{jouve09}, but $R_e$ is about a factor of two smaller.

This simulation is more laminar than some others done with the ASH code \citep[e.g.][]{miesch08,jouve09}, and we acknowledge that this may have an impact on our results.  However, 
the focus of this paper is the fundamental mechanisms by which thin flux tubes interact with global 
convection and mean flows.  The convection simulation possesses all the relevant features necessary to investigate this interaction, including columnar, asymmetric, rotationally-aligned cells at low latitudes (density-stratified banana cells), a rapidly-evolving downflow network at higher latitudes in the upper convection zone, dominated by helical plumes, and a strong, solar-like differential rotation.  The convective turnover timescale is $\tau \sim$ 2\e{6} s, or about 23 days. Our work provides a baseline for future work which can further assess how the details of the convective flow affect the results.    Since even the highest-resolution simulations exhibit similar basic features, we do not expect the essential results to change significantly.  The principle effect we expect at higher Rayleigh and Reynolds numbers is an increase in the random scatter due to stochastic turbulent fluctuations.  Decoherence of large-scale motions and turbulent drag from small-scale motions could also slightly reduce rise times but we expect this effect to be minor since we believe that large-scale, columnar banana-like cells must persist even in highly turbulent parameter regimes in order to provide the requisite Reynolds stresses to account for the solar differential rotation.

\begin{figure}
\epsscale{0.75}
\plotone{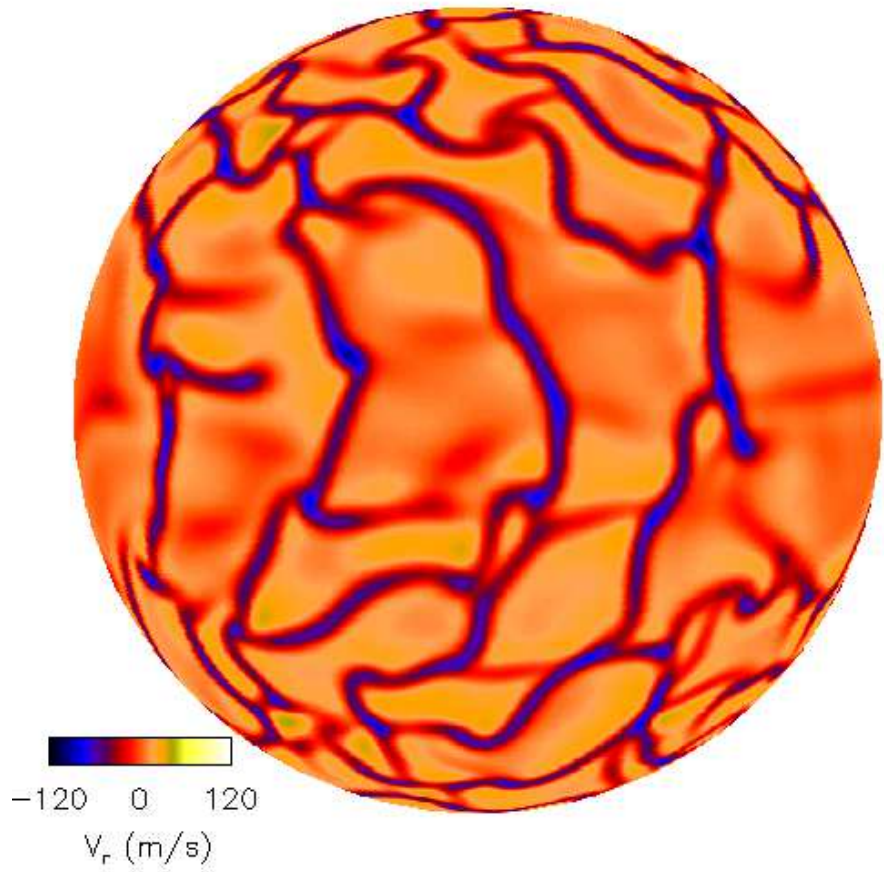}
\caption{Snapshot of convective radial velocity at a depth of 25 Mm below the solar surface. A movie showing the evolution of the radial velocity at this depth for over a period of about 315 days is available in the electronic version of the paper}
\label{fig:vr_sphere}
\end{figure}

\begin{figure}
\epsscale{1.}
\plotone{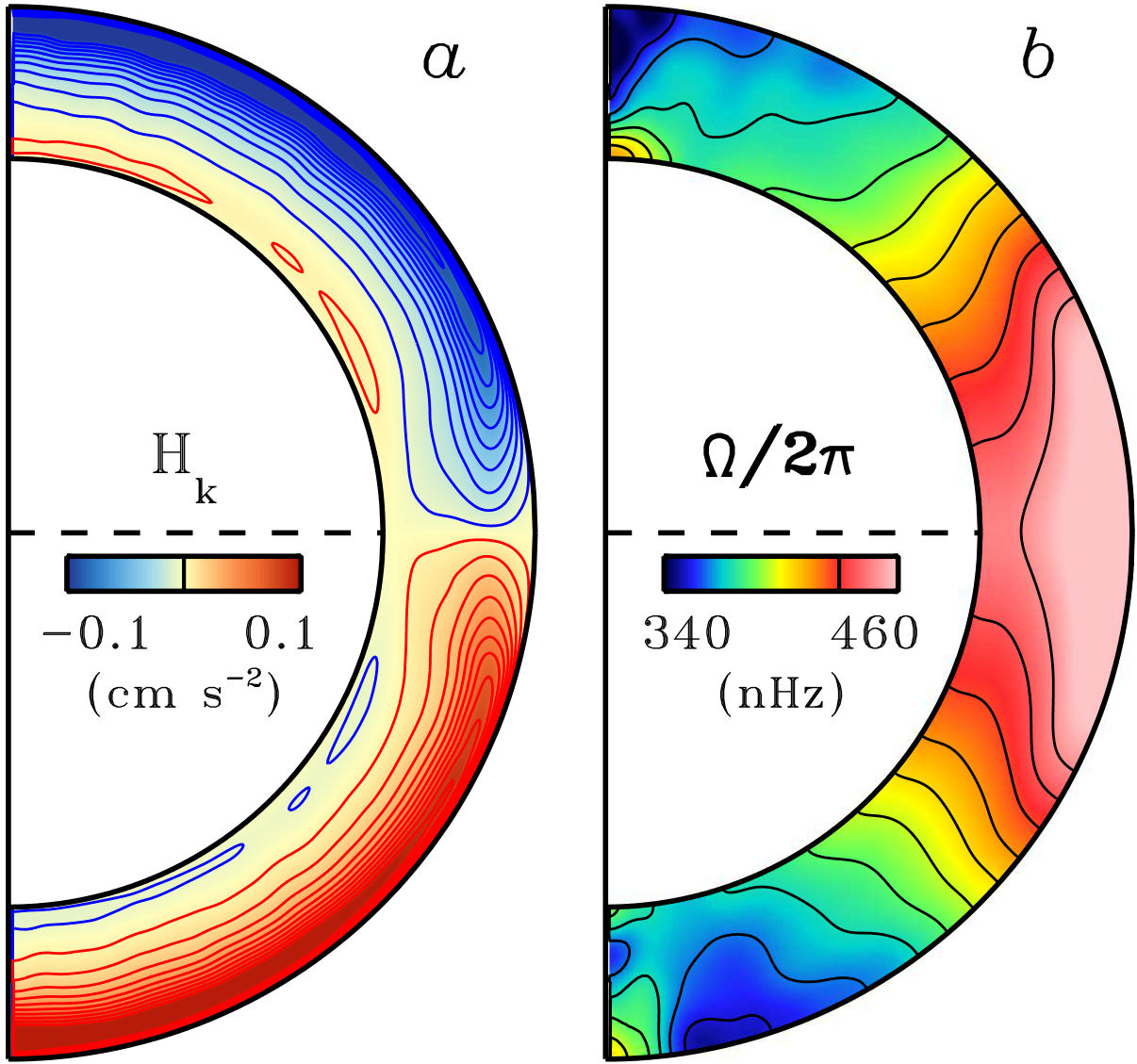}
\caption{($a$) Mean kinetic helicity density $H_k$ and ($b$) angular velocity (with respect to the inertial frame) in the convection simulation, averaged over longitude and time (time interval 1366 days).  Color tables saturate at the values indicated, with extrema ranging from ($a$) -0.122--0.133 cm s$^{-1}$ and ($b$) 326-468 nHz.}
\label{fig:mean_flows}
\end{figure}

Figure \ref{fig:vr_sphere} shows a snapshot of the radial velocity of the giant-cell convection at a depth of $25$ Mm below the solar surface.  An associated movie showing the evolution of the convective flow pattern at this depth for over a period of about 315 days is also available in the electronic journal. The convective flow pattern shows broad upflow cells surrounded by narrow and intense downflow lanes.  The maximum downflow speed in the convective envelope reaches nearly $600$ m/s at a depth of about $86$ Mm below the surface.  Throughout most of the convection zone, the combined influence of the Coriolis force and the density stratification induces anti-cyclonic vorticity in expanding upflows and cyclonic vorticity in contracting downflows.  This yields a mean kinetic helicity density $H_k = \left<\mbox{\boldmath $\omega \cdot v$}\right>$ which is negative in the northern hemisphere and postive in the southern hemisphere (Fig.\ \ref{fig:mean_flows}$a$), where \mbox{\boldmath $\omega$} is the vorticity of the convective flow with velocity \mbox{\boldmath $v$}.  There is a weak sign reversal of $H_k$ in the lower convection zone where downflows expand and recirculate, inducing anti-cyclonic vorticity.  Such a helicity pattern is typical for rotating, compressible convection \citep[e.g.][]{miesch_toomre2009}.

At low latitudes there is a preferential alignment of elongated downflow lanes with the rotation axis, reflecting the presence of so-called ``banana cells'' (Fig.\ \ref{fig:vr_sphere}). These features propagate in a prograde sense relative to polar regions, due in part to the differential rotation and in part to an intrinsic phase drift akin to traveling Rossby waves \citep[e.g.][]{miesch_toomre2009}.  Such structures dominate the convective Reynolds stress, maintaining a strong differential rotation comparable to that inferred from helioseismic inversions.  In particular, the total angular velocity $\Omega/2 \pi$ (with respect to the inertia frame) decreases monotonically from about 470 nHz at the equator to about 330 nHz at the poles and exhibits nearly conical contours at mid latitudes (see Fig. \ref{fig:mean_flows}$b$), as in the solar convection zone \citep{thompson2003}.

\begin{figure}
\epsscale{.5}
\plotone{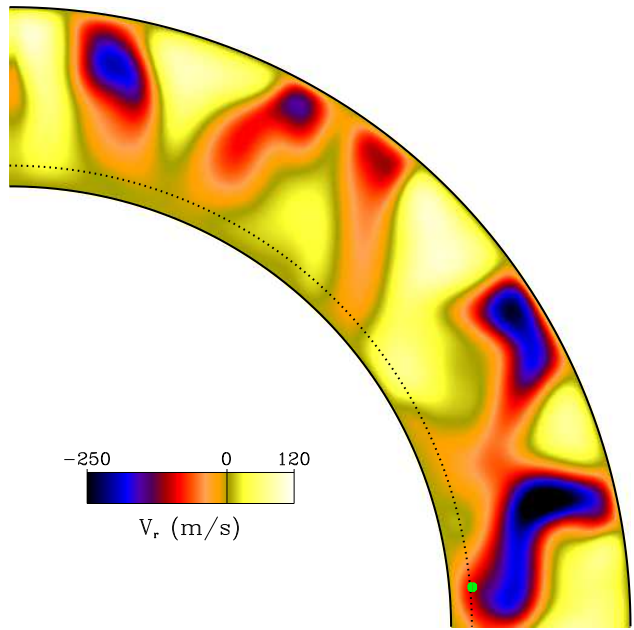}
\caption{Snapshot of a slice of the radial velocity field ($r = 0.69R_{\odot}$ to $0.97R_{\odot}$, $\theta$ from north solar pole to equator) at an arbitrary azimuthal angle $\phi$, with a cross-section of a 15 kG flux tube (green, drawn to scale), at its initial starting radius 6$^{\circ}$ above the equator.  The dotted line represents the base of the convection zone at 5.026\e{10} cm.}
\label{fig:vr_tube}
\end{figure}

As is in \citet{fan_gong2000}, our simulations start with toroidal magnetic flux rings in mechanical equilibrium (neutral buoyancy), located at a radial distance to the center of the Sun $r = r_0 = 5.05 \times 10^{10}$ cm, slightly above the base of the solar convection zone at $r = r_{\rm czb} = 5.026 \times 10^{10}$ cm.   Figure \ref{fig:vr_tube} shows a snapshot of the radial velocity at an arbitrary azimuthal angle $\phi$, with the green dot representing the radius (to scale) of the largest flux tube, which occurs for magnetic field strengths of 15 kG. This figure shows the flux tube at its initial starting position compared to the base of the convection zone.  Note that the convective velocity field is allowed to penetrate into the overshoot region.  To ensure initial neutral buoyancy, the internal temperature of the flux tube is reduced compared to the external temperature.   We consider a range of initial field strengths, with $B_0$ = 15 kG, 30 kG, 40 kG, 50 kG, 60 kG and 100 kG, and initial latitudes $\lambda_0$ ranging from $1^{\circ} $ to $40^{\circ}$ for the toroidal flux ring.  Considering the root mean square (rms) of the convective downflows from the ASH simulation at the base of the convection zone which are on the order of 35 m/s, the equipartition magnetic field is on the order of $B_{eq}\sim5$ kG.  In this case, we are investigating flux tubes on the order of $3-20$ $B_{eq}$.   In all cases, the flux of the tube is constant at $10^{22}$ Mx, on the order of large scale solar active regions.   For this study, we perform two sets of simulations sampling different time ranges of the ASH convective flow at each magnetic field strength for each initial latitude in both the northern and southern hemispheres.  The toroidal ring in neutral buoyancy is perturbed with small undular motions which consist of a superposition of Fourier modes with azimuthal order ranging from $m = 0$ through $m = 8$  with random phase relations.  In this work, the external time dependent 3D convective flow described above also impacts the flux tube through the drag force term, and we study its effect.

%------------------------------------------------------------------------------------------------------------------------------------

\section{Results}
\label{sec:results}

\subsection{Overview of Influence of Convection on Rising Flux Tubes}
\label{sec:overall}
\begin{figure}
\epsscale{1}
\plotone{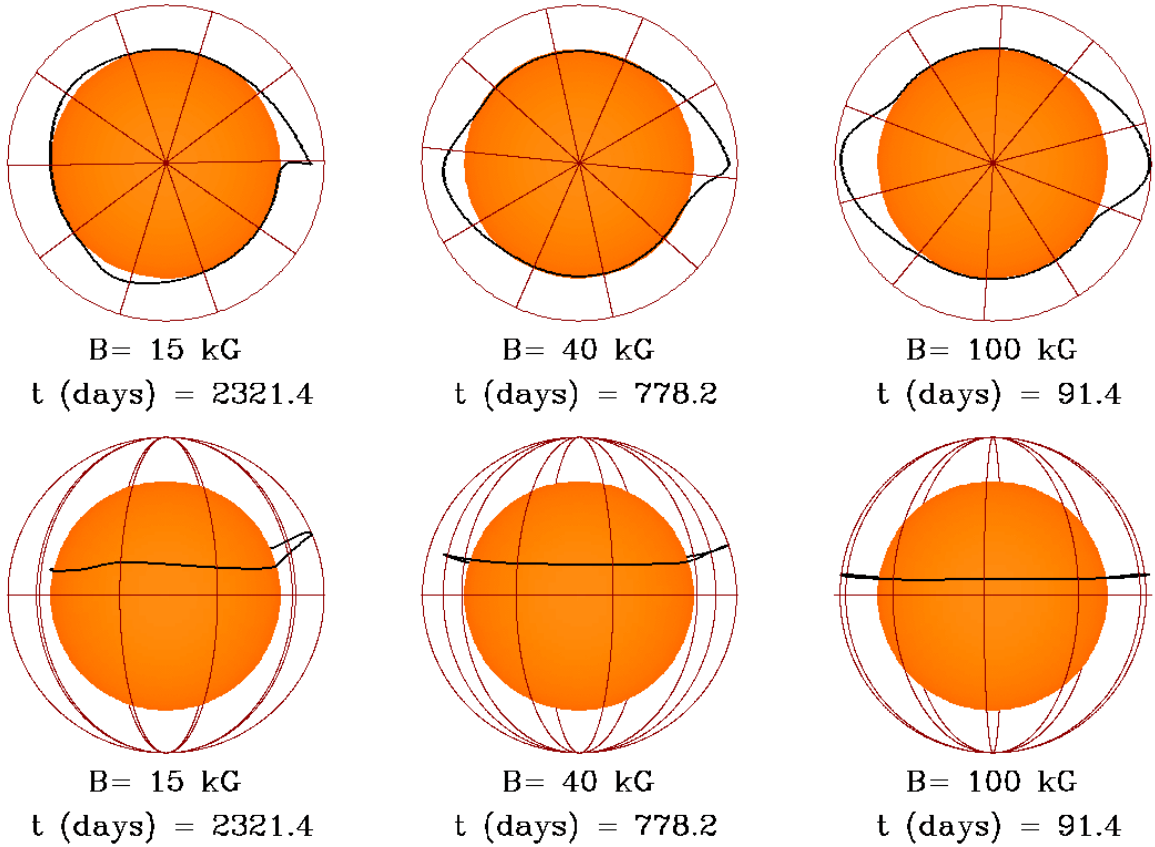}
\caption{Snapshots of the rising flux tube, developed from an initial toroidal flux ring at a latitude of 6$^{\circ}$, at a time when its apex is approaching the top boundary,  with initial magnetic field strengths of 15 kG, 40 kG, and 100 kG respectively, in the absence of convection. The top images show a polar view, whereas the bottom images show an equatorial view.  In all cases, the image has been rotated such that the flux tube apex is on the right, and at the 3 o'clock position if looking down from the north solar pole.  The orange sphere has a radius of 4.9\e{10} cm, which is 0.1\e{10} cm above the base of the rotating solar convective envelope, and 0.15\e{10} cm below the initial position of the flux tubes.}
\label{fig:snapshot_noconv}
\end{figure}

Figure \ref{fig:snapshot_noconv} shows snapshots of the rising flux tube in the absence of convection, developed from an initial toroidal flux ring at a latitude of 6$^{\circ}$, at a time when its apex is approaching the top boundary,  with initial magnetic field strengths of 15 kG, 40 kG, and 100 kG respectively.  The top three images show the flux tube as if looking down on the sun from the north pole, whereas the bottom three images show the flux tube as if looking directly at the equator.  For these cases in the absence of convection, emerging loops develop as a result of the non-linear growth of the magnetic buoyancy instability of the initial toroidal flux tube, and the time for the flux tube to rise to the surface decreases with increasing magnetic field strength. The rise time is 6.4, 2.1, and 0.25 yrs for 15, 40 and 100 kG respectively.  The rise of the loop becomes more radial with increasing initial field strength, with emergence latitudes of 23.3$^{\circ}$, 18.8$^{\circ}$, and 7.8$^{\circ}$ for 15, 40, and 100 kG magnetic fields respectively.  In Figure \ref{fig:snapshot_noconv}, we note the appearance of predominately $m = 2$ modes for magnetic field strengths of 40 kG and 100 kG.  The 15 kG  flux tube shows a predominately $m = 1$ mode, superposed with a very small amplitude $m = 3$ mode.  The results are similar to those in \citet{fan_gong2000}, where the toroidal flux ring develops the undular buoyancy instability \citep[e.g.][]{cali95}, with $m = 1$ and $m = 2$ being the dominant unstable modes, and $\Omega$-shaped rising loops form.      

\begin{figure}
\epsscale{1}
\plotone{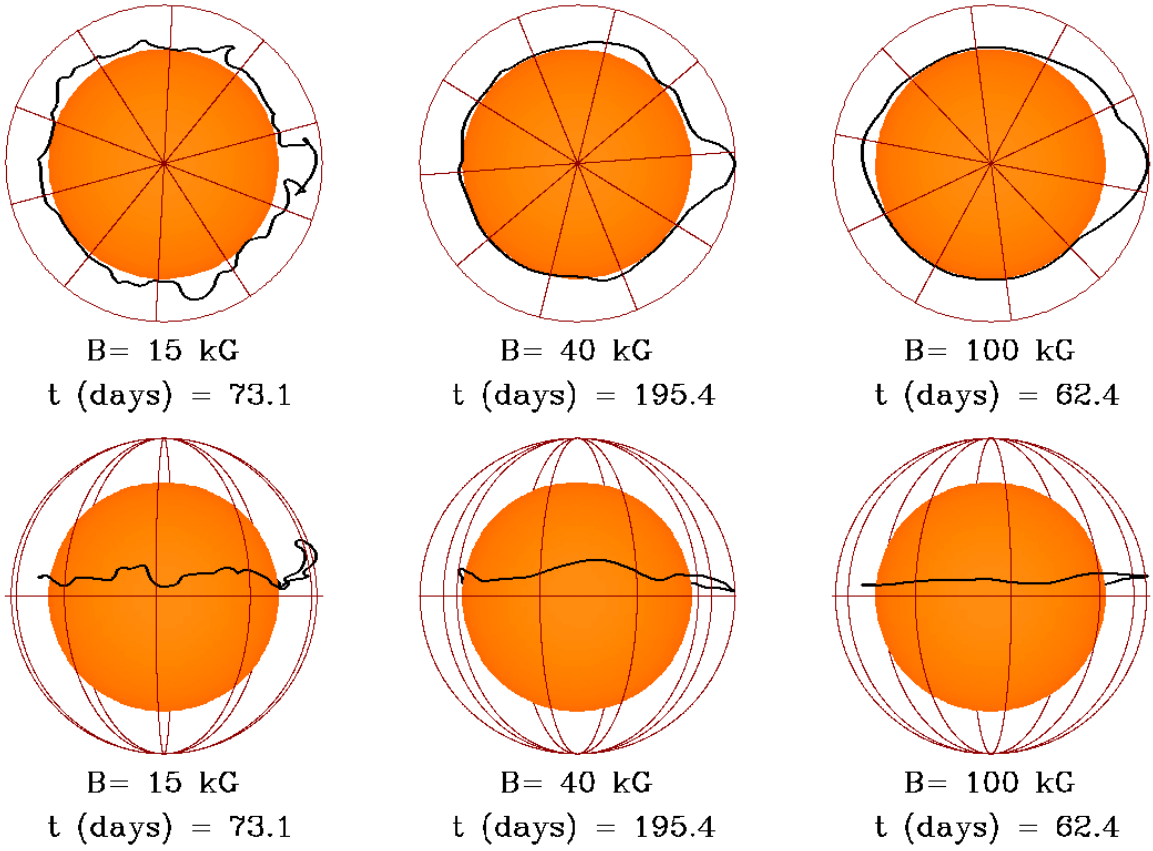}
\caption{Same as Figure \ref{fig:snapshot_noconv}, except the tube is subjected to the external convective flow.}
\label{fig:snapshot_conv}
\end{figure}

Figure \ref{fig:snapshot_conv} shows snapshots of the rising flux tubes with the same initial conditions as Figure \ref{fig:snapshot_noconv}, but with the influence of the external convective flow included. It can be seen that with a low initial field strength of 15 kG, the development of the rising loops are largely controlled by the convective flows.  Convection drastically decreases the rise time from 6.4 yrs to only 0.2 yrs.  On the other hand, at a large magnetic field strength of 100 kG, we find that the development of the rising loops are still mainly controlled by the development of the magnetic buoyancy instability, with the strongest convective downdrafts producing some moderate perturbations to the final emerging loops.  In this case, the presence of convection also shortens the rise time by about one month. At 40 kG, the influence of convection is somewhere in between the above two cases.  Again convection reduces the rise time, from 2.1 yrs to 0.5 yrs in this case. Interestingly, with convection, the flux tube with 40 kG initial field takes longer to rise than either cases with 15 kG or 100 kG.  The predominately $m = 2$ mode for the 100 kG flux tube is still discernible.  However, at lower magnetic field strengths the flux tube evolution is altered by convective flows such that a dominant mode can no longer be established.                

\begin{figure}
\epsscale{.6}
\plotone{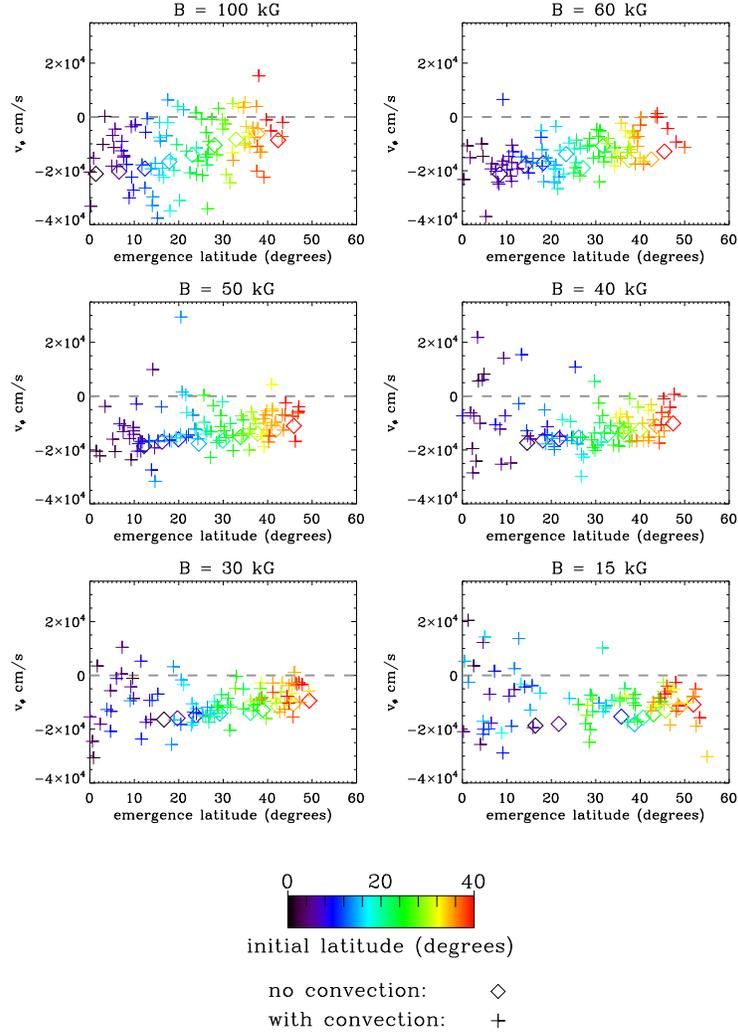}
\caption{Average velocity in the azimuthal direction $\phi$ of the mass elements in the legs of the emerging flux loop once it has reached the top of the simulation domain.  These values are plotted for initial magnetic field strengths of 100kG, 60 kG, 50 kG, 40 kG, 30 kG, and 15 kG, respectively, in the presence of convection (plus signs), and the absence of convection (diamonds).}
\label{fig:vphi}
\end{figure}

The differential rotation of the convective velocity field does have an influence on the evolution of the rising flux tubes, especially at low magnetic field strengths.  Figure \ref{fig:vphi} plots the average velocity in the azimuthal direction $\phi$ of the mass elements in the legs of the emerging flux loop once it has reached the top of the simulation domain.  We find that without convection, the average velocity of the mass elements of the emerging loop is always negative. This occurs because, as the tube rises, the conservation of angular momentum drives a retrograde motion of particles in the tube, producing a negative velocity in the $\phi$ direction.  At a large magnetic field strength of 100 kG, the averaged azimuthal velocity of the emerging loop in the presence of convection is found to be centered around the azimuthal velocity for the tubes without convection.  With decreasing magnetic fields, the azimuthal velocity in the presence of convection becomes preferentially faster than in the case without convection.  This indicates that differential rotation is attempting to push the elements of the flux tube prograde compared to the case without convection, especially at low magnetic field strengths where tubes are more susceptible to convective flows.             

\begin{figure}
\epsscale{.5}
\plotone{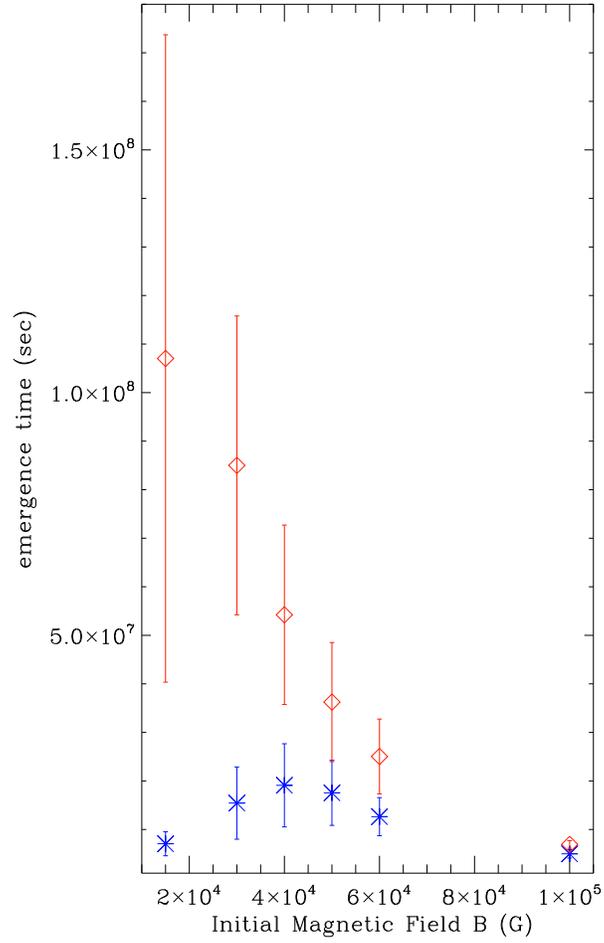}
\caption{Average time for the flux tube to rise through the convection zone for magnetic fields of 100 kG, 60 kG, 50 kG, 40 kG, 30 kG, and 15 kG in the absence of convection (red diamond points), and in the presence of convection (blue star points).  Bars represent the variation in the rise time for different starting latitudes of the flux tube.}
\label{fig:rise_times}
\end{figure}

Figure \ref{fig:rise_times} summarizes how long it takes for an $\Omega$-shaped loop to emerge at the top of the simulation domain.  Without convection, the time for the flux tube to emerge decreases with increasing field strength.  This agrees with \citet{moreno83}, where they find the same trend for flux tubes given sinusoidal perturbations.  Also, the range of variation in rise time between tubes initiating from different latitudes (as represented by the size of the error bar centered on the mean in Fig. \ref{fig:rise_times}) decreases with increasing initial magnetic field.  When the influence of convection is included, a different trend emerges, and the rise times are no longer monotonic with increasing magnetic field strength.  In all cases, the rise time is reduced with the addition of convection, especially for the cases with weaker fields.  The rise time becomes the longest for mid-field strength cases (i.e. 40 kG).  At these mid-field strengths, the effects due to the magnetic buoyancy and the average convective downflows influencing the rise of the flux tube are of similar magnitudes, and work against each other, keeping the flux tube in the mid-convection zone for a longer period of time than at other magnetic field strengths.  These effects will be discussed more in detail in the following paragraphs.  For all the field strengths investigated, convection is found to reduce the variation of rise times for flux tubes with different initial latitudes, as can be seen in the size of the error bars compared to the case without convection.

To understand the importance of convection, we compare the magnitude of the magnetic buoyancy force with that of the drag force from the external convective flows, following \citet{fan03}. For the drag force to dominate buoyancy:
\begin{equation}
 \frac{C_{D} \rho_{e} v_{e}^2}{\pi a} > \frac{B^2}{8\pi H_{p}},
\label{eq:compare}
\end{equation}
where $H_{p}$ is the local pressure scale height, and $a$ is the flux tube radius $a=(\Phi/\pi B)^{1/2}$, with flux of the tube $\Phi = 10^{22}$ Mx.    The pressure scale height ranges from 5.6\e9 cm at the base of the convection zone, to 6.9\e8 cm at the top of the simulation domain.  Also, the radius of the flux tube spans from 4.6\e8 cm for an initial magnetic field strength of 15 kG, to 1.8\e8 cm for 100 kG.  For an order of magnitude estimate of the magnetic buoyancy here, we have assumed thermal equilibrium between the flux tube and the external fluid. Assuming $2C_{D}/\pi$ is of order 1, then equation (\ref{eq:compare}) simplifies to
\begin{equation}
v_{e} > v_{a} \bigg( \frac{a}{H_{p}} \bigg)^{1/2}.
\label{eq:compare2}
\end{equation}
i.e. in order for convection to dominate, the convective flow speed $v_e$ needs to be greater than the Alfv\'en speed  $v_a=B/(4\pi\rho_{e})^{1/2}$ multiplied by $(a/H_p)^{1/2}$.

\begin{figure}
\epsscale{.75}
\plotone{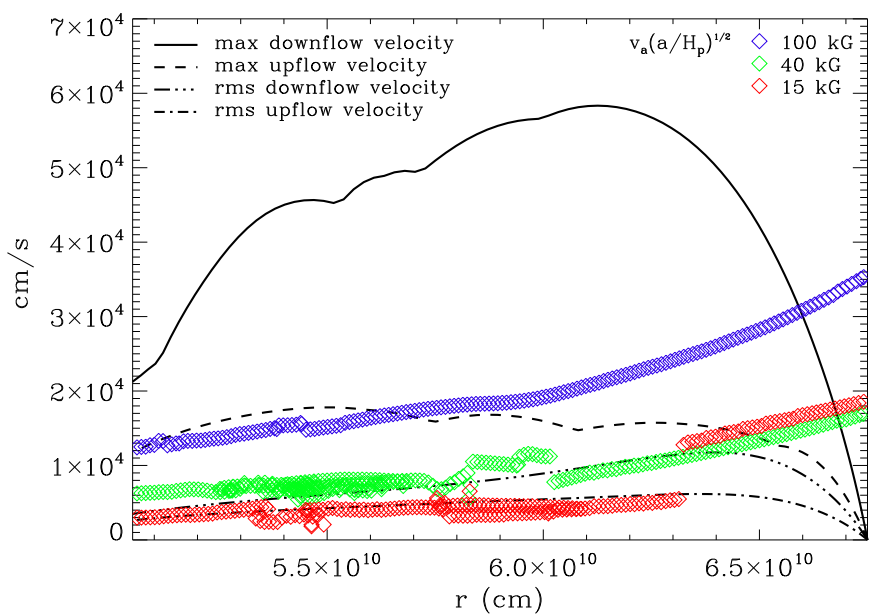}
\caption{The peak down-flow, peak up-flow, the root mean square (rms) of the down-flow, and the rms of the up-flow of the convection velocity field in each constant r surface as a function of $r$.  Also plotted are the right hand side of Eq. \ref{eq:compare2} at the apex of each of the the flux tubes shown in Figure \ref{fig:snapshot_conv} with different initial field strengths, as it traverses the convection zone.}
\label{fig:compare_forces}
\end{figure}

\begin{figure}
\epsscale{.75}
\plotone{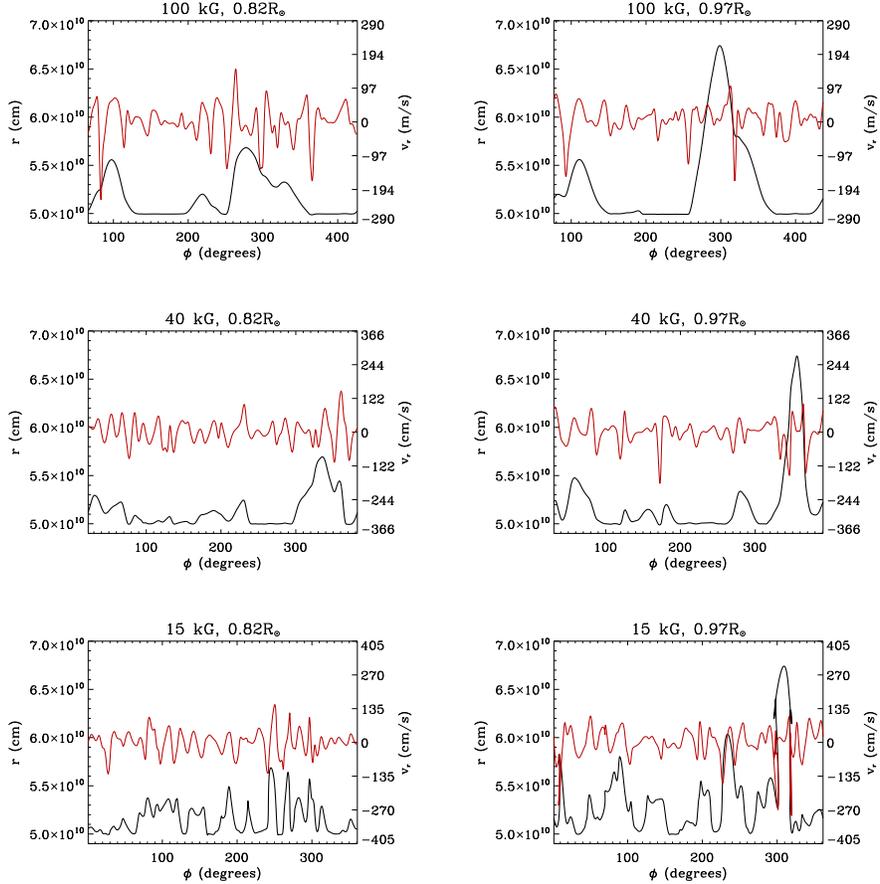}
\caption{Flux tube radial distance from center of the Sun $r$ (black line), plotted with the external radial velocity experienced by the flux tube at the height $r$ of the flux tube segment (red line),  both as functions of the azimuthal angle $\phi$.  Figures on the left are for an instance when the apex height is $r=0.82R_{\odot}$, and on the right are for the last time step when the apex has reached the top of the simulation domain at $r=0.97R_{\odot}$.  These snapshots show all the azimuthal angle values for the flux tube, with all tubes originating at 6$^{\circ}$ latitude.}
\label{fig:vr}
\end{figure}

In Figure \ref{fig:compare_forces}, we have plotted as a function of depth the peak downflow, peak upflow, the root mean square (rms) of the downflow, and the rms of the upflow of the convection velocity field at each constant $r$ surface.  In comparison, we have also plotted the right hand side of Eq. \ref{eq:compare2} evaluated at the apex (portion of flux tube with largest $r$ value) of each of the flux tubes shown in Figure \ref{fig:snapshot_conv} with different initial field strengths, as it traverses the convection zone.  It can be seen that for the 100 kG case, only the largest convective downflows are strong enough to impact the rise of the flux tube, whereas at 15 kG, even the rms of the convective flows, both upflows and downflows, can significantly affect the rise of the tube.  In the case of 40 kG, all of the convective velocity field except for the rms upflows contribute to the development of the rising loops. These explain the general behavior we see in Figure \ref{fig:snapshot_conv}.  Note in Figure \ref{fig:compare_forces} that there is a concentration of points for the 40 kG (green) and 15 kG (red) cases.  This is due to the fact that the flux tube is continually buffeted by convection, with strong downflows pushing down some apices such that new apices will take over for portions of the flux tube evolution.  Often, the apices are also aided by boosts from convective upflows.  Eventually, one loop becomes buoyant enough such that it will finally emerge at the top of the simulation domain.  There is a jump in the 15 kG curve that occurs near 6.3\e{10} cm because one loop is boosted by a convective upflow and becomes significantly more buoyant than the previous apex.  It is clear from Figure \ref{fig:compare_forces}, that downflows in the convective velocity field dominate in amplitude, however their spatial extent is small compared to the upflows as indicated by the narrow downflow lanes shown in Figure \ref{fig:vr_sphere}.  Strong downflows can pin the flux tubes to the base of the convection zone at the beginning of the simulation faster than the Fourier mode perturbations can in the case without convection.   Especially at low magnetic field strengths where the tube is highly susceptible to convection deformation, many portions of the tube will become anchored as compared to the case without convection.  Also, rising loops can be significantly boosted by broad upflows such that they emerge at the top of the simulation domain much faster than in the case without convection, where only buoyancy aids in driving the tube to the surface.  Convection can also enhance buoyancy instabilities by introducing finite-amplitude perturbations and subsequent gravity induced draining of fluid from the flux tube apex.  Figure \ref{fig:vr} shows snapshots (one individual time instance) of the radial distance $r$ of the flux tube from Sun center as a function of the azimuthal angle $\phi$ (black lines), as well as the external radial velocity experienced by the tube at the height $r$ of each tube segment (red lines).  These snapshots are for the same flux tubes as used in Figures \ref{fig:snapshot_conv} and \ref{fig:compare_forces}, with the left column showing the snapshot at a time when the apex of the tube has reached 0.82$R_{\odot}$, and the right column showing the snapshot for the last time step when the flux tube apex has reached the top of the simulation domain at 0.97$R_{\odot}$.  From this figure, it is evident that at large magnetic field strengths, only the strongest downflows can perturb the tube.  However, at small magnetic field strengths, all flows are capable of deforming the flux tube.   

\subsection{Latitude of emergence and tilt angles}
\label{sec:emandtilt}

\begin{figure}
\epsscale{.75}
\plotone{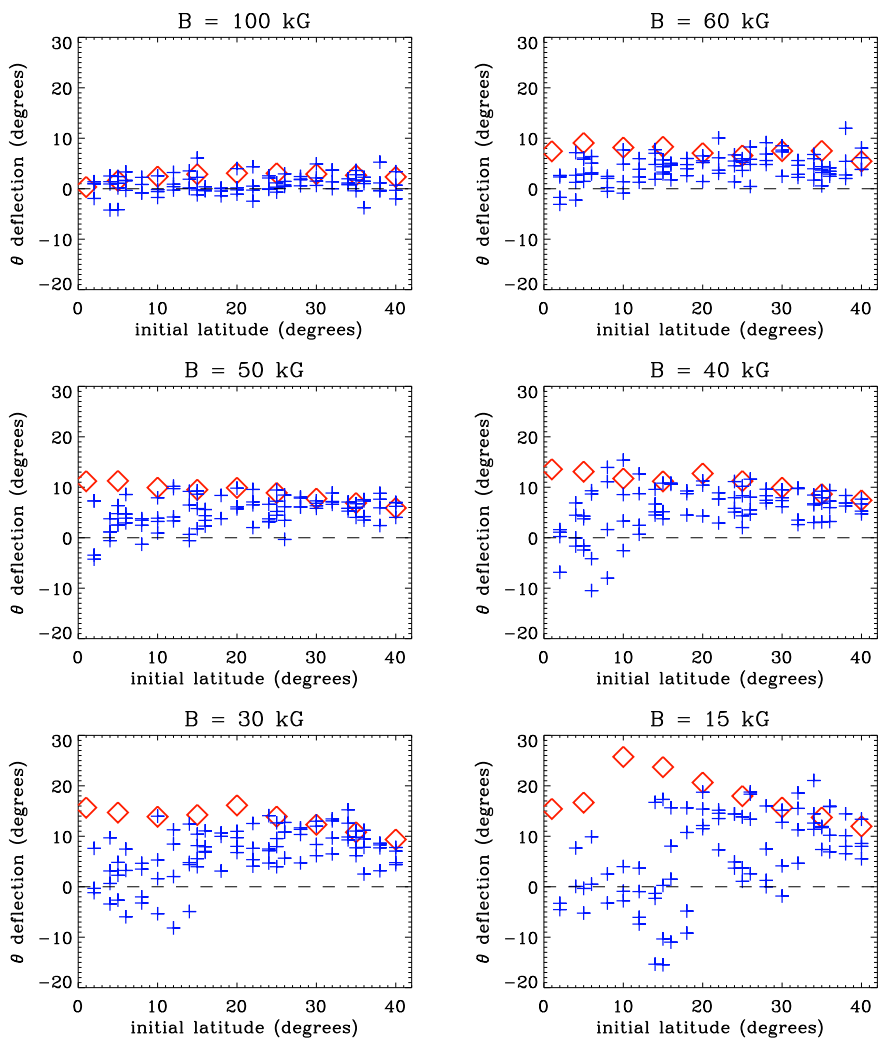}
\caption{Latitudinal deflection (emergence latitude minus initial latitude) of flux tube apex as a function of initial latitude for initial tube field strengths of 100 kG, 60 kG, 50 kG, 40 kG, 30 kG, and 15 kG in the absence of convection (red diamond points), and in the presence of convection (blue plus sign points).}
\label{diff}
\end{figure}

In the absence of convection, we find that for weaker initial field strengths, there is a low latitude zone devoid of flux emergence due to the poleward deflection of the rising tube by the Coriolis force \citep[e.g.][]{choud87, cali95}.  This is shown in Figure \ref{diff}, where we have plotted the deflection of the flux tube in latitude (emergence latitude minus initial latitude) as a function of the flux tube initial latitude.  The red diamonds (blue plus signs) show the deflection of the flux tubes without (with) the influence of convection.  This low latitude zone void of flux emergence in the case without convection shrinks as the initial magnetic field strength is increased.  This is because at larger initial magnetic field strengths, the buoyancy force overpowers the Coriolis force such that tubes rise more radially.  However, when the flux tubes are subjected to the convective flow, we find that they are able to emerge at lower latitudes near the equator, even for low field strength cases.  In some cases, the convective velocity field is able to push the flux tube such that its apex will emerge closer to the equator than where the tube initially started.  This is indicated in Figure \ref{diff} by negative plus sign points.  We note that with convection, the poleward deflection of the flux tubes is reduced for the majority of cases.  Flux tubes subjected to turbulent convection spend less time in the convection zone than the flux tubes without convection (see Fig. \ref{fig:rise_times}).  As such, the Coriolis force will have less time to deflect these flux tubes to higher latitudes.  Also, increased anchoring of weaker flux tubes by convection reduces the fraction of the tube that moves away from the rotation axes, and therefore reduces the effect of the Coriolis force on the tube.                       

\begin{figure}
\epsscale{.6}
\plotone{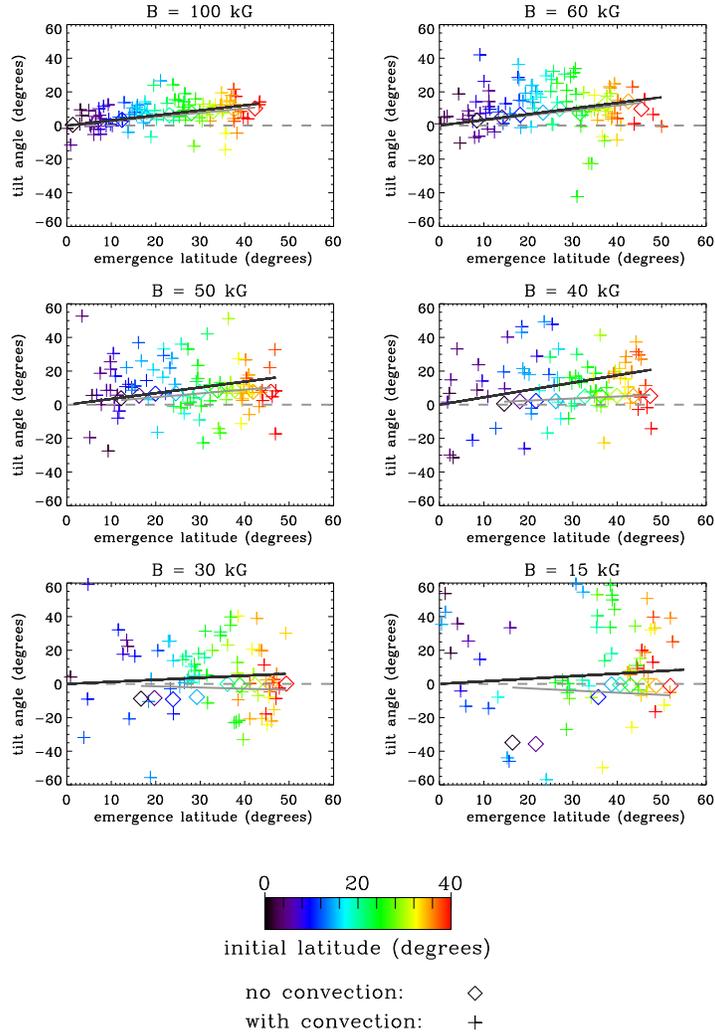}
\caption{Tilt angle as a function of emergence latitude for initial magnetic field strengths of 100 kG, 60 kG, 50 kG, 40 kG, 30 kG, and 15 kG respectively, and for cases with (plus signs) and without (diamond points) the influence of convection. The gray line is the least squares fit for flux tubes in the absence of convection.  The best fit line slopes with uncertainties for field strengths of 100 kG, 60 kG, 50 kG, 40 kG, 30 kG, and 15 kG are $0.25\pm0.01$, $0.30\pm0.02$, $0.22\pm0.02$, $0.12\pm0.01$, $-0.07\pm0.05$, and $-0.13\pm0.13$. The black line is the fit to the cases subjected to the convective flow, with best fit line slopes for fields strengths of 100 kG, 60 kG, 50 kG, 40 kG, 30 kG, and 15 kG of $0.30\pm0.08$, $0.34\pm0.05$, $0.34\pm0.12$, $0.44\pm0.14$, $0.12\pm0.18$, and $0.15\pm0.21$.  A color bar indicates the original starting latitude of the flux tube.  Dashed lines indicate 0$^{\circ}$.  }
\label{fig:tilt_angles}
\end{figure}

In most cases without convection, the variation of the emergence latitude with initial latitude is smooth, except for the 15 kG flux tube, where there is an abrupt 14$^{\circ}$ jump in emergence latitude between the tubes originating at 5$^{\circ}$ and 10$^{\circ}$ (see diamond points in the bottom right panel of Fig. \ref{diff} and Fig. \ref{fig:tilt_angles}).  At this magnetic field strength in the absence of convection, flux tubes originating at 5$^{\circ}$ and 1$^{\circ}$ have troughs which become anchored to the overshoot region below the base of the convection zone.  However for the same tubes originating from 8$^{\circ}$ and above, the rising tubes do not become anchored.  As such, the tube floats upward as a whole, and experiences a relatively greater deflection towards the pole and hence emerges at a higher latitude compared to the anchored cases.  In these cases, the $m=1$ to $m=3$ modes grow slower than the $m=0$ mode.  This corresponds to a poleward slip of the tube as a whole.  Thereafter, modes on the order of $m = 1$ to $m = 3$ begin to develop larger amplitudes, allowing loops to develop, though the troughs do not anchor at the base of the convection zone.  With convection, at this same magnetic field strength, all of the flux tubes have troughs which become anchored because convective downflows pin them to the bottom of the convection zone from the beginning of the simulation.  We note that the flux tubes tend to anchor (when they anchor at all) around 5.0\e{10} cm in both the cases with and without convection.  At this depth, the sub-adiabaticity is $\delta=$-1.3\e{-4}.  The flux tubes all originate at 5.05\e{10} cm, 2.8\e{8} cm above the base of the convection zone, and have a super-adiabaticity here of $\delta=$3.2\e{-8}, or very nearly zero.  Reducing the sub-adiabaticity would reduce the chances for anchoring of the portions of the flux tube that are perturbed into the overshoot region.   Increasing the sub-adiabaticity would help anchor these flux tubes if they were perturbed into the overshoot region and did not remain anchored at a reduced sub-adiabaticity.  However, we find that the tubes which do not anchor in this study are never perturbed into the overshoot region.  Instabilities of higher order with large enough amplitude to perturb the tube into the overshoot region do not develop fast enough to overcome the $m=0$ mode.   
 
Figure \ref{fig:tilt_angles} shows the tilt angle of the emerging flux loops as a function of the emergence latitude, for flux tubes with a range of initial latitudes (from $1^{\circ}$ to $40^{\circ}$ as indicated by the color of the points). The different panels in the Figure show the results for initial magnetic field strengths of 15 kG, 30 kG, 40 kG, 50 kG, 60 kG, and 100 kG respectively.  In the plots, results with (without) the influence of convection are shown as plus signs (diamonds). 
  
It is well known that bipolar active regions on the sun tend to form with the leading polarity (in the direction of solar rotation) being slightly closer to the equator than the following polarity, and thus showing a tilt angle for the line connecting the following to the leading polarities with respect to the east west direction, as described by Joy's Law \citep[e.g.][]{hale19, espuig10}.  On average, the tilt angle of the active region increases with its latitude. In Figure \ref{fig:tilt_angles} the tilt angle is computed as the angle between the tangent vector at the apex of the emerging loop and the local east-west direction. We define a positive sign of tilt as a clockwise (anti-clockwise) rotation of the tangent vector away from the east-west direction in the northern (southern) hemisphere, consistent with the direction of the observed mean tilt of active regions.  If on the other hand the magnitude of the tilt angle exceeds 90$^{\circ}$, then it means that there is an anti-Hale polarity arrangement for the active region.  

It is important to note that we do not consider the effect of magnetic field line twist on the evolution of flux tubes.  \citet{fan08} considers the effect of twist on the tilt angle of emerging flux tubes for a 3D MHD simulation in a local Cartesian geometry without an external velocity field.  They find that in order for emerging flux tubes of highly super-equipartition field strength on the order of $10^{5}$ G  to have tilts consistent with Joy's Law,  the initial twist rate of the flux tube needs to be smaller than about a half of that required for the tube to rise cohesively.  However, in the case of flux tubes in a convecting box with equipartition magnetic field strengths on the order of $10^{4}$ G, it is found that convective downflows dominate over magnetic buoyancy effects, and the evolution of the tube is no longer sensitive to the initial twist of the tube \citep{fan03}.  In this paper, we are primarily concerned with how convection alters the overall evolution of the rising flux tubes, and the thin flux tube model used here does not address the effect of the field line twist.            

\citet{espuig10} analyzed Mount Wilson and Kodaikanal sunspot group tilt angle and latitude data spanning solar cycles 15-21.  In order to re-derive Joy's Law, they performed a fit for the equation: $\alpha=m\lambda$ to the data, where $\alpha$, $m$, and $\lambda$ represent the tilt angle, the slope, and the latitude respectively.  The fit is forced to go through zero because no tilt is expected for equatorial sunspot groups.  They obtained a slope of $0.26\pm0.05$ and $0.28\pm0.06$ for Mount Wilson and Kodaikanal data, respectively. 

In Figure \ref{fig:tilt_angles} we show the same kind of linear least squares fit for both the simulated tilt angles with and without the influence of the convective flow. The slope of the best fit lines with uncertainties for 100 kG, 60 kG, 50 kG, 40 kG, 30 kG, and 15 kG without convection are respectively: $0.25\pm0.01$, $0.30\pm0.02$, $0.22\pm0.02$, $0.12\pm0.01$, $-0.07\pm0.05$, and $-0.13\pm0.13$. Thus the slopes of the best fit lines fall within the range found by \citet{espuig10} only for flux tubes with initial field strengths of 100 kG, 60 kG, and 50 kG.  For weaker initial field strengths (below about 40 kG) the emerging loops begin to show negative tilt angles, opposite to the sign of the active region mean tilts.  This occurs because flow along the flux tube at the loop apex changes from diverging into converging as it enters the upper layers of the convection zone \citep[e.g.][]{cali95, fan_fisher1996}, and the Coriolis force acting on the converging flow drives a tilt of the wrong sign. With the influence of the convective flow included, the slopes of the best linear fits to the tilt angles all increase due to the systematic effect of the kinetic helicity (see Fig.  \ref{fig:mean_flows}$a$) associated with the rotating convective flows, which tends to drive rotation at the apex of the loops in the direction consistent with the sign of the observed active region mean tilts.  The slope of the best fit lines with convection for initial magnetic field strengths of 100 kG, 60 kG, 50 kG, 40 kG, 30 kG, and 15 kG are respectively: $0.30\pm0.08$, $0.34\pm0.05$, $0.34\pm0.12$, $0.44\pm0.14$, $0.12\pm0.18$, and $0.15\pm0.21$.  As a result, the tilt angles of emerging loops with initial field strengths of 100 kG,  60 kG, 50 kG, and 40 kG are all consistent with Joy's Law.  These field strengths all show a positive Joy's Law trend, given the uncertainties in the fitted slopes.  However, at lower field strengths, the tilt angles show large random scatters produced by convection, and do not show a significant systematic dependence on latitude as described by Joy's law.  The uncertainty of the slope also tends to increase with decreasing magnetic field, reflecting the random scatter produced by convection.  While the slope of the best fit lines for 30 kG and 15 kG are still positive, the uncertainties are too large to report a definitive positive Joy's Law trend for these field strengths.  There is a larger than expected uncertainty on the slope for 100 kG.  This occurs because there is an outlier, which can not be seen in Figure \ref{fig:tilt_angles} as it falls outside the region graphed.  The outlier corresponds to a flux tube which originates in the southern hemisphere, but emerges in the northern hemisphere such that the emergence of the tube would result in an anti-Hale polarity arrangement for the active region.       

\subsection{Asymmetry in Field Strength}
\label{sec:bee}
Another well-known observed asymmetry is in the morphology of the leading and the following polarities of an active region, where the leading polarity flux tends to be concentrated into a well formed sunspot, whereas the following polarity flux tends to appear more fragmented and dispersed \citep[e.g.][]{bray_loughhead1979}. The thin flux tube simulations of \citet{fan93} showed that the preceding leg of an emerging magnetic flux loop has a stronger magnetic field than the following leg as a result of the differential stretching of the rising loop due to the Coriolis force.  This gives an explanation for the observed more coherent and less fragmented morphology for the leading polarity flux in an active region. However subsequent simulations \citep[e.g.][]{cali95, fan_fisher1996} using the mechanical equilibrium initial state (which is more physical) as opposed to the temperature equilibrium as used by \citet{fan93}, found that the leading leg of the emerging loop has a stronger magnetic field than the following only for flux tubes with an initial field strength that is below about 60 kG.  For flux tubes with higher initial field strengths, the field strength asymmetry reverses at the top of the loop.

\begin{figure}
\epsscale{.75}
\plotone{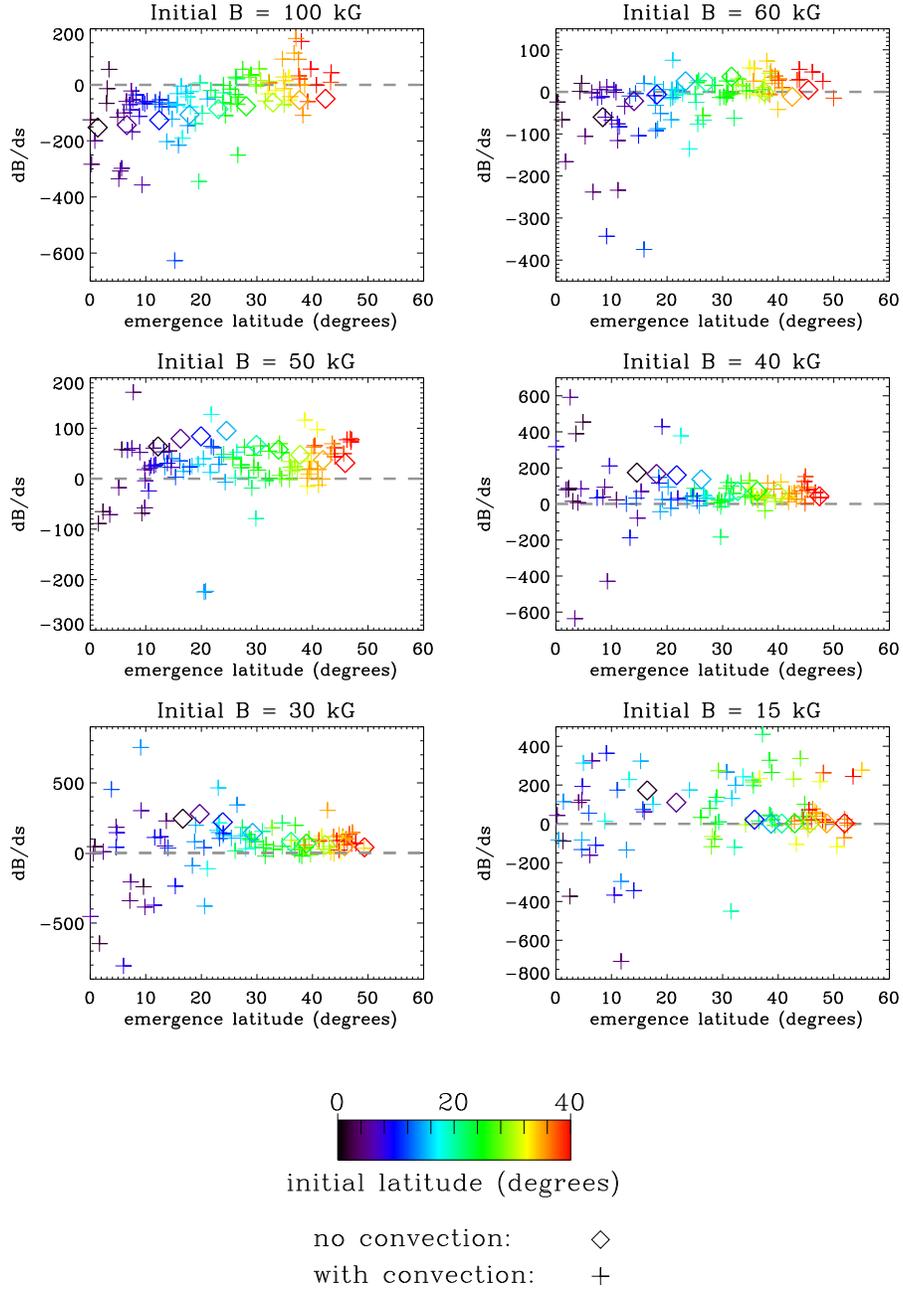}
\caption{$dB/ds$ at the apex of the emerging flux loop as a function of emergence latitude for tubes with an initial field strength of 100 kG, 60 kG, 50 kG, 40 kG, 30 kG, and 15 kG respective, for cases with (plus signs) and without (diamond points) the influence of convection. A color bar indicates the original starting latitude of the flux tube.  Note that the y-axes are not the same for every plot.}
\label{fig:dbds}
\end{figure}

Here we investigate this magnetic field asymmetry by calculating $dB/ds$ at the apex of the emerging flux loop as shown in Figure \ref{fig:dbds}.  If $dB/ds$ is greater (less) than zero, then the leading (following) leg has a stronger magnetic field. As can be seen in Figure \ref{fig:dbds},  at 50 kG and below, the majority of the emerging loops show stronger field in the leading leg than the following leg in the presence of convection.  However, at 100 kG, with convection, only some of the emerging flux loops at higher latitudes have stronger field in the preceding leg.  At 60 kG there are about equal number of positive and negative $dB/ds$ cases. Thus with convection, loops with initial field $\leq 50$ kG tend to emerge with the appropriate magnetic field asymmetry (i.e. with a stronger field in the preceding leg and therefore can be expected to result in an emerging active region with a more coherent leading polarity).

\subsection{Asymmetry of Inclination}
\label{sec:inc}
When looking down on the emerging loops from the pole, it is apparent that there is an asymmetry in the inclination of the leading and following legs with respect to the vertical direction, as is evident in the cases without convection shown in Figure \ref{fig:snapshot_noconv}.  The following leg tends to have a steeper slope than the leading leg.  This asymmetry is caused by the Coriolis force and the conservation of angular momentum as the tube rises through the convection zone \citep{moreno94, cali95, cali98}, and provides an explanation for the apparent asymmetric east-west proper motions of the two polarities of an emerging active region.  As the tube rises above a constant r surface, inclination of the legs of the loop cause an apparent more rapid motion of the leading polarity spot away from the emerging region as compared to the motion of the following polarity spots. 

\begin{figure}
\epsscale{.75}
\plotone{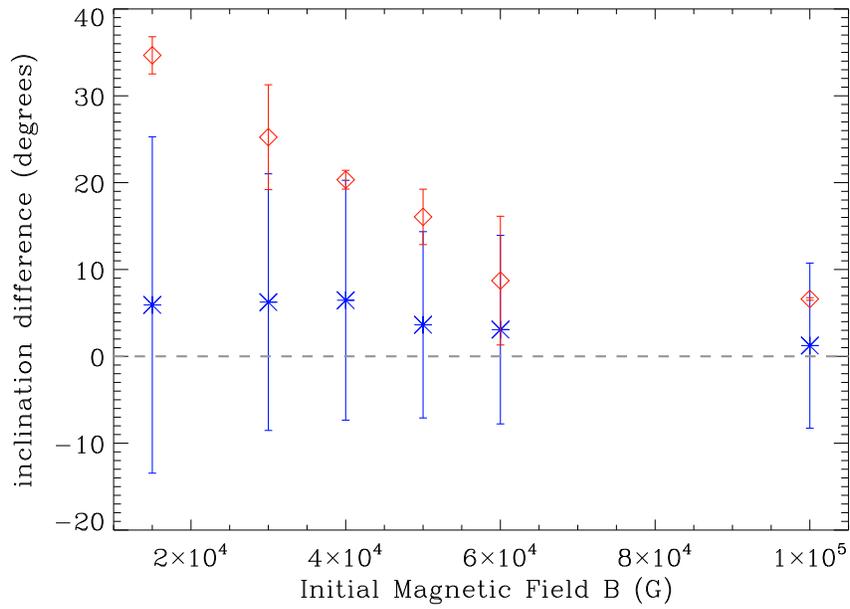}
\caption{Average difference of the inclination angles (see description in the text) between the leading and the following sides of the emerging loop for initial magnetic field strengths of 100 kG, 60 kG, 50 kG, 40 kG, 30 kG, and 15  kG respectively, for cases with (blue star points) and without (red diamond points) the influence of convection. Bars represent the variation in inclination angle difference for different starting latitudes of the flux tube.}
\label{fig:inclin_diff}
\end{figure}

Here we quantify the steepness of each leg of the emerging loop by calculating what we call the inclination angle.  To do this, first we find the portion of the emerging loop which is concave downward.  Then we find the best fit line for each leg from the flux tube apex in the concave down portion.  The angle between the best fit line for each leg and $- {\hat {\bf r}}$ is the inclination angle.  The smaller this angle, the steeper the slope of the leg.  In Figure \ref{fig:inclin_diff} we have plotted the average difference of the inclination angles (leading leg minus the following leg).  Only the average inclination difference for each magnetic field strength is plotted because we find that the inclination difference does not vary systematically with latitude.  The bars are the standard deviation, showing the spread in inclination angle difference for each magnetic field strength.  This spread is due primarily to convective effects.  Positive inclination difference means the following leg is steeper, consistent with the results of \citet{moreno94} and \citet{cali95, cali98}.  We find that at all field strengths, the majority of the emerging loops develop a steeper slope for the following leg.  The inclination angle differences are overall reduced with the inclusion of convection.   

It is worthwhile to note that convection aids in bringing the footpoints of the rising flux loops closer together, most notably in the 15 kG and 40 kG cases as can be seen in Figure \ref{fig:snapshot_noconv} and Figure \ref{fig:snapshot_conv}.  This can also be observed in Figure \ref{fig:vr}, where strong convective downflows pin portions of the tube down, creating many buoyant loops with very close footpoints at 15 kG.

%-----------------------------------------------------------------------------------------------------------------------------------

\section{Discussion}
\label{sec:discuss}
In subjecting the thin flux tube model to a turbulent solar convective velocity field computed separately from a 3D global convection simulation in a rotating spherical shell representing the solar convective envelope, we study how convection can influence the development and emergence of solar active region flux tubes. Although idealized, the thin flux tube approximation allows us to investigate flux emergence at moderate to strong field strengths (15 - 100 kG) in localized flux concentrations under perfect flux frozen-in conditions.  We find that as the field strength of the initial toroidal flux ring placed near the bottom of the solar convective envelop is increased from $15$ kG to $100$ kG, the development and the evolution of the emerging flux loops change from being convection dominated to magnetic buoyancy dominated.  At 15 kG initial field strength, the development of the emerging flux loops is largely controlled by the convective flows.  Both the average up-flows and down-flows can significantly affect the rise of the flux tube. On the other hand, at 100 kG initial field strength, the development of the emerging loops are largely controlled by the growth of the magnetic buoyancy instability and only the strongest convective downdrafts can significantly impact the rising tubes.

With the inclusion of the convective flows, the previous issue of 15 kG - 30 kG flux tubes being significantly deflected poleward by the Coriolis force during their rise is resolved.  Loops can now emerge at low latitudes all the way to the equator.  The main reasons for this are the anchoring of the emerging flux loops by the convective downdrafts and also the faster rise of the loops propelled by the upflows. Convection is also found to produce random scatters in the tilts of the emerging loops, especially for the weaker field strength cases where the scatters are greater.  In addition, because the convective flow in the rotating spherical envelope shows a mean kinetic helicity that is negative (positive) in the northern (southern) hemisphere, it on average tends to drive clockwise (anti-clockwise) tilts for the rising loops in the northern (southern) hemisphere, consistent with the sign of the active region mean tilts.  As a result, the inclusion of convection tends to increase the slope of the linear Joy's Law fit of the tilt angles of the emerging loops as a function of emerging latitudes.  Flux tubes with initial field $\gsim$ 40 kG and $\lsim$ 100 kG, are found to produce a  mean tilt angle dependence on latitude that is consistent with the observed Joy's Law. However the effect of the kinetic helicity in the convective flow is not sufficient, on average,  to correct the tendency for loops to develop tilts of the wrong sign (opposite to that of the observed active region mean tilt) for tubes with initial field strength $< 40$ kG.

Similar to \citet{cali95} and \citet{fan_fisher1996}, we find that at or below 50 kG, the leading leg of the emerging loop tends to have a larger magnetic field than the following, which may provide an explanation for the observed better cohesion of the leading polarity of an emerging active region compared to the following polarity. This trend of asymmetry in field strength reverses for tubes with initial field $> 60 $kG.  Also we find that the inclusion of convection does not change the general trend of the asymmetry in the geometry of the emerging loop found in previous studies \citep[e.g.][]{cali95}, where the leading leg tends to be more horizontally inclined than the following, which would provide an explanation for the apparent asymmetric east-west proper motions of the two polarities of an emerging active region.  This trend of geometric asymmetry is not dependent on the magnetic field strength.

These results combined suggest that in order for the emerging loops to be consistent with the observed properties of solar active regions, the best range of field strengths for toroidal magnetic field at the bottom of the convection zone is $\sim$ 40 - 50 kG.  Above this range, the magnetic field is strong enough to resist convection such that the flux tube properties are not changed significantly from the case without convection, but the magnetic field asymmetry of the emerging loops tend to be of the wrong sense.  Below this range, convection dominates the development and evolution of the emerging loops.  The loops are able to emerge at low latitudes, but their tilt angles show very large scatter.  Increasing the number of simulations run to collect more data may result in a reduction of the uncertainty on the slope of the Joy's Law best fit lines, such that 15 kG and 30 kG flux tubes may also show a positive Joy's Law trend.                 

The main limitation of the thin flux tube approach is that the model assumes the tube remains cohesive, and does not address the dynamic effect of the internal twist of the flux tube. Also, as the ASH convective flow is computed separately, the calculations do not address how the rising flux tube might affect the surrounding convective flow. Full 3D spherical-shell MHD simulations \citep[e.g.][]{jouve09} are necessary to address the above questions, but it is still beyond the current computational capabilities to adequately resolve an active region scale flux tube in a global convection simulation.  It is important to note that these results are for flux tubes which have reached 21 Mm below the solar surface.  Even though these flux tubes still must traverse a small portion of the convection zone and encounter the solar surface shear layer, we feel that the behavior of the flux tubes at this depth may give a good representation of the large scale pattern of flux emergence at the solar surface.  This paper serves as an initial study to understand how active region scale flux tubes of various strengths are affected by convective flows as they rise through the solar convective envelope.

\section{Acknowledgements}
This work is supported in part by NASA SHP grant NNX10AB81G to the National Center for Atmospheric Research (NCAR).  NCAR is sponsored by the National Science Foundation.  We would also like to thank the anonymous referee for the very thorough report and helpful comments.  

%-----------------------------------------------------------------------------------------------------------------------------------

%-----------------------------------------------------------------------------------------------------------------------------------

\clearpage

%figure 1

%figure 2

%figure 3

%figure 4

%figure 5

%figure 6

%figure 7

%figure 8

%figure 9

%figure 10

%figure 11

%figure 12

%figure 13

%figure 14

%-----------------------------------------------------------------------------------------------------------------------------------

\end{document}